\documentclass[%
 aip,
 jmp,
 amsmath,amssymb,
preprint,%
]{revtex4-1}

\usepackage[dvips]{color}
\usepackage{graphicx}
\usepackage{dcolumn}
\usepackage{bm}
\usepackage{ulem}

\begin{document}
\title{\color{blue}{Lasing in planar semiconductor diodes}} %Title of paper
\author{Giorgio De Simoni}
\affiliation{Scuola Normale Superiore, Laboratorio NEST, I-56127 Pisa, Italy}
\author{Lukas Mahler}
\affiliation{NEST, Istituto Nanoscienze-CNR and Scuola Normale Superiore, I-56127 Pisa, Italy}
\author{Vincenzo Piazza}
\affiliation{Center for Nanotechnology Innovation @ NEST, Istituto Italiano di Tecnologia, I-56127 Pisa, Italy}
\author{Alessandro Tredicucci}
\affiliation{NEST, Istituto Nanoscienze-CNR and Scuola Normale Superiore, I-56127 Pisa, Italy}
\author{Christine A. Nicoll}
\affiliation{Cavendish Laboratory, University of Cambridge, Cambridge CB3 0HE, United Kingdom}
\author{Harvey E. Beere}
\affiliation{Cavendish Laboratory, University of Cambridge, Cambridge CB3 0HE, United Kingdom}
\author{David A. Ritchie}
\affiliation{Cavendish Laboratory, University of Cambridge, Cambridge CB3 0HE, United Kingdom}
\author{Fabio Beltram}
\affiliation{Scuola Normale Superiore, Laboratorio NEST, I-56127 Pisa, Italy}
%\date{\today}   

\begin{abstract}
We present a planar laser diode based on a simple fabrication scheme compatible with virtually any geometry accessible by standard semiconductor lithography technique. We show that our lasers exhibit $\sim$1 GHz -3dB-modulation-bandwidth already in this prototypical implementation. Directions for a significant speed increase are discussed.
\end{abstract}

\maketitle
Semiconductor light-emitting diodes (LEDs) and laser diodes are key components of today's light-generation technologies. They provided the easiest link between electronics and photonics and, thanks to their low power consumption, long lifetime and high reliability, they spread out of the telecommunication field and are now used for countless applications and systems. A peculiar LED class is that of planar (or lateral) light-emitting diodes (LLEDs): in these devices the $n$ and the $p$ part do not overlap along the semiconductor growth axis, but are hosted by the same layer, whose thickness can reach the limit of few nanometers in the case of 2-dimensional electron and hole gases (2DEG and 2DHG) confined in a heterostructure quantum well (QW)\cite{Cecchini2003,Hosey2004,Kaestner2002}. The planar geometry allows LLEDs to take advantage of the modulation-doping technique to reduce non-radiative recombination channels and makes them suitable for applications that strictly require two-dimensionality, \textit{e.g.} integration with high-electron-mobility transistors or surface-acoustic-wave-driven devices\cite{DeSimoni2009}. At the same time, LLEDs are very appealing in terms of frequency-modulation bandwidth\cite{Cecchini2005}. Indeed, junction size can be much smaller than in vertical diodes and this can lead to a sizable reduction of total device capacitance. The latter characteristic is even more interesting if applied to a laser diode: a conventional quantum-well (QW) or multiple-QW laser-diode exploits quantum confinement to benefit from the advantages brought by the reduced dimensionality in terms of joint-density of states. However, since the carrier transport is perpendicular to the QW(s), the device remains essentially 3-dimensional. The very few realizations of planar laser diodes presented so far were based on the amphoteric properties of silicon as a dopant on GaAs substrate\cite{North1999,Ryzhii2002}. This approach allowed to demonstrate the feasibility of a LLED, but, since the device had to be ridge-shaped and parallel to a specific crystal axis, there was no degree of freedom in the orientation and design of the optical cavity. Also, for such devices, no evidence of good performance in terms of modulation bandwidth was ever shown. In the following, we present a prototype of LLED that exploits a simple fabrication scheme compatible with virtually any geometry allowed by standard semiconductor lithography techniques. Thanks to a Bragg grating defined in the waveguide, our devices are also able to emit both from the facets and from the surface and we show $\sim$1 GHz -3dB-modulation-bandwidth, which can be much improved merely by reducing laser dimensions.

\begin{figure}[h!]
\centering
\includegraphics[width=8.5cm]{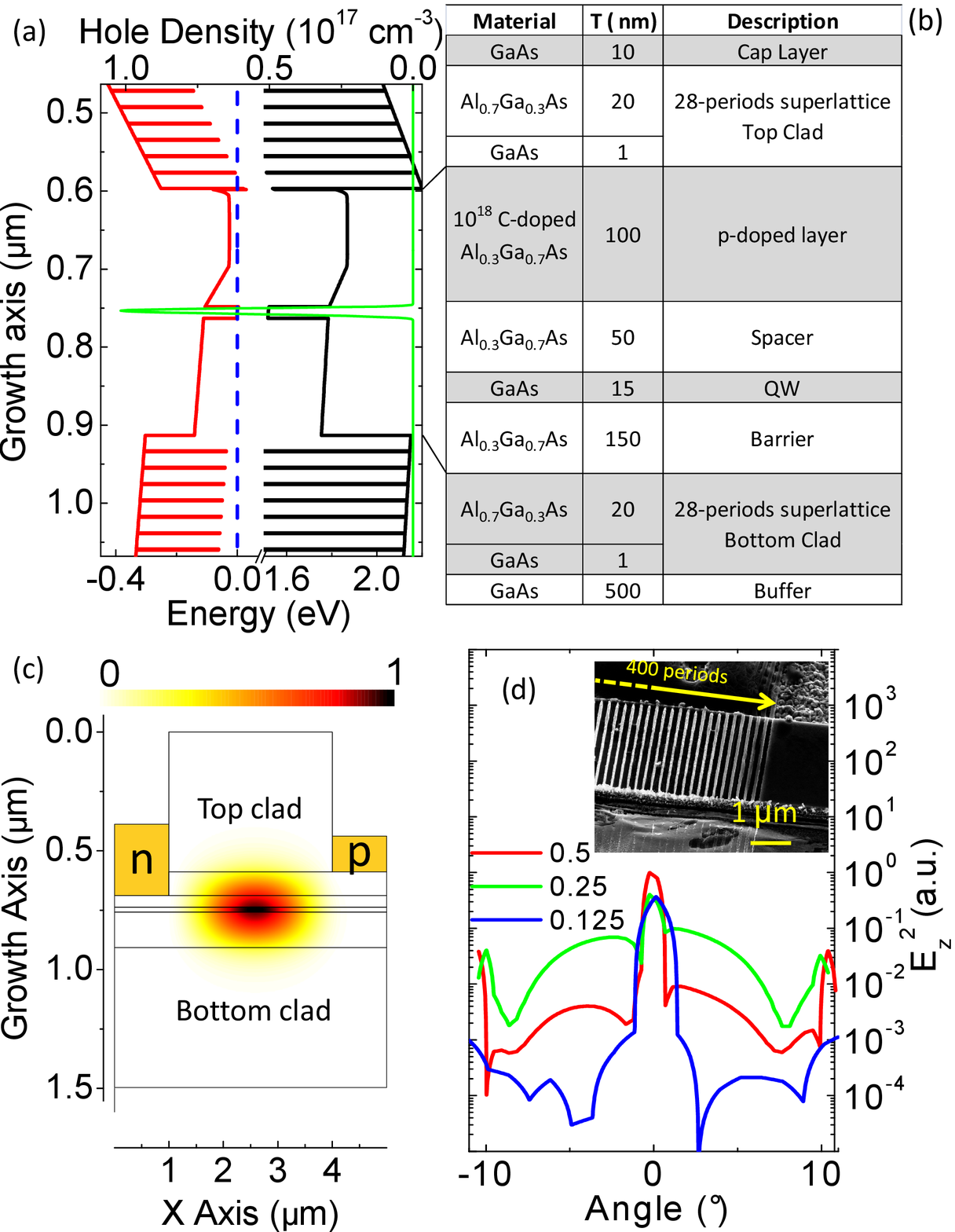}
\caption{(a): Top of valence (red) and bottom of conduction (black) band energy level with respect to the Fermi energy (blue dashed line) as a function of the depth along the heterostructure growth axis. The hole density (green line) is also reported. Bands and charge profiles were obtained by a self-consistent Schr\"oedinger-Poisson calculation. (b): Heterostructure layer composition. Thickness (column labeled as T) and description of the layers are reported. (c): A device section scheme. The yellow rectangles represent the $n$- (left) and $p$-type (right) Ohmic contacts. The total electromagnetic energy density of the single-lobe TE mode with frequency equal to 3.66$\cdot$10$^{14}$ Hz is color-surface plotted in energy arbitrary units. The edges of the devices and of the heterostructure layers are indicated for clarity. (d): Far field angular distribution for different etch depths (0.5, 0.25 and 0.125 of UCL total thickness) of the grating. The simulated ridge has infinite width. Inset: A scanning-electron-microscope image of the device A. Part of the extraction grating is visible.}
\label{fig1}
\end{figure}

Our devices are based on a carbon-modulation-doped GaAs/Al$_{x}$Ga$_{1-x}$As heterostructure hosting a 2DHG (Poisson-Schr\"odinger self-consistently calculated charge density is 2$\cdot$10$^{10}$ cm$^{-2}$ at 4 K) confined in a 15 nm-wide GaAs QW (the valence/conduction band and charge density calculations are plotted in Fig.~\ref{fig1}a). Optical confinement along the growth axis is provided by a planar dielectric waveguide whose cladding layers are formed by two GaAs/Al$_{0.7}$Ga$_{0.3}$As superlattices (1 nm/20 nm for a total thickness of 588 nm). The GaAs layers were inserted as smoothing layers to avoid roughness at the interface of 70\% AlGaAs layer and the core region, which consists of the QW and two 150 nm-thick Al$_{0.3}$Ga$_{0.7}$As spacers. Lateral optical confinement is obtained by removing the upper cladding layer (UCL) without the need to remove core or bottom cladding layers. This geometry makes possible to define a lateral waveguide without removing or even depleting the QW from carriers. Electrical access to the 2DHG is obtained by locally removing the UCL and thermally evaporating in the same region a Au/Zn/Au (5/50/100 nm) annealed Ohmic contact. The $n$-part of the devices is fabricated following a similar procedure, but, in addition to the UCL, the C-doped layer is also removed in order to locally deplete the QW from holes. Electrons are then induced by thermal evaporation of a Ni/AuGe/Ni/Au (10/200/10/100 nm) annealed Ohmic contact. During the annealing procedure, the evaporated Ge diffuses into the semiconductor and acts as a donor creating an electron gas in the volume underlying the contact. In this way, a planar $p$-$n$ junction is formed inside the QW at the interface between the $n$-doped volume and the 2DHG. This approach, sketched in Fig. \ref{fig1}b, makes possible to provide lateral optical confinement and fabricate the diode in the same processing steps. We highlight that this fabrication scheme does not strictly require the device to be ridge-shaped, but is limited only by technical details of the lithographic procedures. More specifically, the same approach can be applied to curved, disk- or ring-shaped laser resonators.

A finite-element-method (FEM) simulation allowed us to calculate the spatial distribution of the electro-magnetic wave-mode in a section of an infinite-length 3-$\mu$m-wide ridge-shaped device at fixed frequency equal to 3.66$\cdot$10$^{14}$ Hz, corresponding to the frequency of the inter-band transition in GaAs. Simulations confirmed the existence of several guided modes. In particular the guide can sustain single-, double-, and triple-lobe modes for both TE and TM polarizations. Single-lobe TE mode electromagnetic energy density spatial distribution is reported in Fig. \ref{fig1}b. While a narrower ridge waveguide can reduce the number of sustained modes, for our sample, the calculated depletion region for a LLED fabricated with this approach is $\sim$1 $\mu$m, which constitutes a lower boundary that must be taken into account for the ridge width. A further simulation step allowed us to design a Bragg grating which supports both edge and surface emission modes by coupling laser radiation in the vertical direction  through second order diffraction. The simulated grating is constituted by a 242 nm periodicity sequence of 50\%-duty-cycle etched/unetched strips in the UCL perpendicular to the ridge. The chosen periodicity corresponds to the wavelength of the inter-band transition in GaAs, properly rescaled to take into account the effective refractive index of the calculated modes supported by the waveguide. The total length of the grating was 96.8 $\mu$m (400 periods) and this was located in the center of the 2 $\mu m$-wide ridge. Several simulations were performed at different etching depth values. Far-field calculations (by the Stratton-Chu formula) for some etching depths are reported in Fig. \ref{fig1} and confirm the vertical extraction of the radiation by means of the grating.

Several devices were fabricated following the described protocol. The UCL layer was etched by an inductively-coupled reactive-ion process to ensure an almost-vertical profile of the ridge side walls. Longitudinal optical feedback was provided by cleaving the samples perpendicularly to the ridge in order to establish a Fabry-Perot resonator. In this article we present results obtained with two devices: one (device A) is 6 mm long, the second (device B) 3 mm. Devices were cooled in a He-flow cold-finger cryostat with optical access and radio-frequency coaxial waveguides in order to test the device modulation-bandwidth performance from room temperature down to 4 K.

\begin{figure}[htb]
	\centering
	\includegraphics[width=8.5cm]{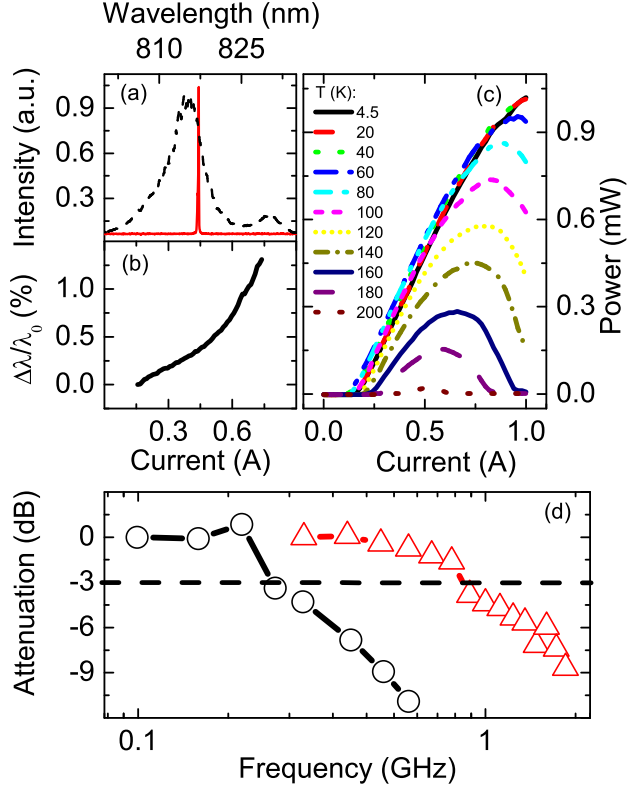}
	\caption{(a) In-plane spectra of device A, below (black curve) and above (red curve) laser threshold. The former corresponds to a current equal to 80 mA; the latter to a current of 250 mA and was filtered by a 10$^5$ neutral density filter to avoid saturation of the collection setup. (b) Spectral evolution of device A laser line with current: the shift of the peak is weighted with the threshold laser wavelength (816 nm). (c) Light-current characteristics of device A at several temperatures. (d) Attenuation of light-intensity oscillations as a function of current-modulation frequency normalized to the low-frequency value. Black and red line respectively refer to device A and B.}
\label{fig2}
\end{figure}

Preliminary voltage-current characterizations showed the expected rectifying behavior with 1.5 V conduction threshold at 4 K.  Electroluminescence spectra were also collected from both the surface and the facets of the devices. The electroluminescence spectrum of device A at 4 K and with 80 mA forward current is reported in Fig. \ref{fig2}a (black line). It exhibits a main electroluminescence peak, centered at $\sim$815 nm, due to electron-hole recombination in the QW. A smaller peak due to electrons recombining in carbon defects was also present at 830 nm. With higher injected current, continuous wave (CW) laser emission from the facets of the devices was observed in a wide range of temperatures. Figure \ref{fig2}a reports the laser spectrum of device A at 4 K and 250 mA current (red line). In order to allow a quantitative comparison with the electroluminescence spectrum, the laser spectrum was rescaled down by a factor of 5$\cdot$10$^5$. The laser line is centered at 816 nm. For injection currents above $\sim$0.2 A, it was observed to shift towards longer wavelengths (830 nm at 1 A). This effect can be attributed to electron-hole recombination below the band-edge due to a Franz-Keldysh-like \cite{HaugH;Koch2004} tunneling-assisted process. This behavior is due to the high electric field ($>$10$^6$ V/m) across the ridge and allows us to tune the emission wavelength by simply changing the bias. Moreover by reconstructing the luminescence spatial profile across the ridge (measurements not reported) we observed that the electric field established between the contacts also strongly affects the emission position: at lower biases (below threshold), the luminescence comes mainly from the edge of $n$-contact due to the electrons being injected into the 2DHG. In the lasing regime, the high electron injection leads to a progressive depletion of the hole gas and determines a luminescence signal to arise from the edge of the $p$-contact. This behavior is explained by the picture of both electrons and holes being injected from the Ohmic contacts into the QW and drifting toward the contact of the opposite species. Due to the high electric field (higher than $\sim$10$^3$ V/m) and thanks to the high-quality material, the electron and hole drift velocities are high enough to let them to travel across the ridge without recombining (i. e.  the radiative recombination time is higher than the ridge-crossing time) and let  them accumulate at the interfaces between the QW and the contacts (due to residual Schottky barriers) where, finally, radiative recombination occurs. This regime is somehow equivalent to ambipolar transport in an intrinsic QW, in which both electrons and holes are provided exclusively by the injection from the Ohmic contacts. 

Our setup did not allow us to fully resolve the internal structure of the laser peak, which appears to be resulting from, at least, two lasing modes. Indeed polarization-resolved spectra showed that it is mainly TE-polarized, with a 20\% of total intensity in the TM component. While best performance in terms of emitted power (up to 1 mW with 1 A injected current) was observed at cryogenic temperatures, CW-laser radiation was detectable up to 200 K. A temperature-dependent light-current characterization of device A is reported in Fig. \ref{fig2}c.

Device modulation bandwidth was probed at 4K by a photon-counting setup that yielded the time-resolved intensity of the light collected by a 300-ps bandwidth avalanche-photodiode. Both devices A and B were biased by a dc voltage (V$_{dc}$) chosen in order to inject $\sim$200 mA across the junction. A small RF signal (in the range between -20 dBm and 0 dBm) was added to V$_{dc}$. Moreover, the devices were 50$\Omega$-terminated in order to avoid reflection of the RF signal. The attenuation of the light-intensity oscillations with respect to the low-frequency case are reported in Fig. \ref{fig2}d as a function of the RF-signal frequency for both devices. Device A and B exhibit 250 MHz and 880 MHz $f_{-3dB}$ modulation bandwidth, respectively. The better performance of device B is due to its smaller size that results in a smaller capacitance, which is roughly expected to scale linearly with the ridge length. For this reason we believe these results to be very promising with respect to further improvement of speed performance of planar laser devices.

\begin{figure}[htb]
\centering
\includegraphics[width=8.5cm]{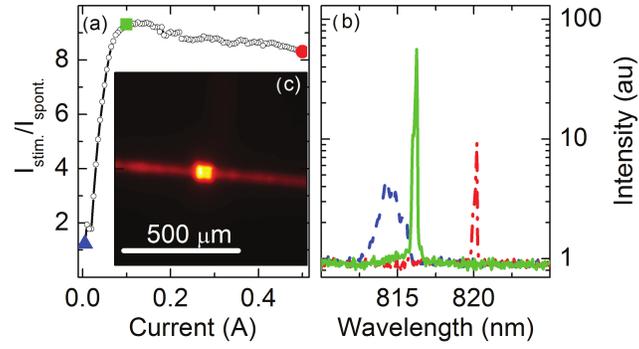}
\caption{a): Ratio between spontaneous and stimulated vertical emission as a function of current. Inset: CCD image of the device vertical emission with an injection current equal to 100 mA. b): Vertical emission spectra at 50 mA (blue), 100 mA (green), 500 mA (red).}
\label{fig3}
\end{figure}

Following this characterization, a grating -with the simulated geometry- was fabricated on device A by means of electron-beam lithography and a 294 nm-deep (corresponding to one half of the UCL total thickness) inductively-coupled reactive-ion etch. Vertical emission was demonstrated at 4 K, by focusing the surface-emitted light on a charge-coupled device (CCD). An image collected by the CCD at 100 mA injection current is reported in Fig. \ref{fig3}c: an intense emission spot is located in correspondence of the grating region, while a spontaneous emission background is emitted from the whole ridge. The ratio of vertical stimulated to spontaneous emission (I$_{stim}$/I$_{spont}$) was estimated by calculating the ratio between the average value of the CCD-pixel intensity inside and outside the grating region. Fig. \ref{fig3}a contains a plot of I$_{stim}$/I$_{spont}$ as a function of the driving current. The emission from the grating shows a steep increase from 0 to 100 mA, then the ratio saturates at a value around 9. A spectral analysis of the vertical emission (reported in Fig. \ref{fig3} showed that the extraction efficiency of the grating is linked to the spectral evolution of the laser as a function of current. In particular the emission is maximized when the laser line is centered at 817 nm. 

In conclusion we have demonstrated a planar laser diode based on a building scheme compatible with several cavity geometries. In addition to in-plane emission, laser surface emission was demonstrated by defining a distributed Bragg grating based extraction region on the ridge surface. Moreover our devices exhibit modulation bandwidths close to 1 GHz and we foresee the possibility of a further increase of their speed performance.

G.D.S would like to acknowledge A. Pitanti and M. S. Vitiello for the useful discussions.
%\bibliography{all}
%

\clearpage
\clearpage
\clearpage
\end{document}